%% file: main.tex
\documentclass[11pt]{article}
\usepackage[T1]{fontenc}
\usepackage{lmodern}
\usepackage[margin=1in]{geometry}
\usepackage{amsmath,amssymb,amsthm,microtype}
\usepackage{booktabs,tabularx}
\usepackage[numbers,sort&compress]{natbib}
\usepackage[hidelinks]{hyperref}
\usepackage{enumitem}
\usepackage{xcolor}
\usepackage{parskip}
\usepackage[linesnumbered,ruled,vlined]{algorithm2e}
\usepackage[normalem]{ulem}
\usepackage{placeins}

\hypersetup{
    colorlinks=true,
    linkcolor=blue,
    linktoc=section
}
\newtheoremstyle{reductionstyle}
  {\parskip}   
  {\parskip}   
  {\itshape}   
  {}           
  {\bfseries}  
  {.}          
  {0.5em}      
  {}

\theoremstyle{reductionstyle}
\newtheorem{reduction}{Reduction}

\theoremstyle{reductionstyle}
\newtheorem{Transformation}{Instance Transformation}

\setlist{itemsep=3pt,topsep=4pt}
\newtheorem{theorem}{Theorem}[section]
\newtheorem{lemma}[theorem]{Lemma}
\newtheorem{proposition}[theorem]{Proposition}
\newtheorem{corollary}[theorem]{Corollary}

\newtheorem{remark}[theorem]{Remark}
\theoremstyle{definition}
\newtheorem{definition}[theorem]{Definition}
\newcommand{\OPT}{\operatorname{OPT}}
\newcommand{\Tok}{\operatorname{Tok}}

\newcommand{\calB}{{\mathcal{B}}}
\newcommand{\calC}{{\mathcal{C}}}
\newcommand{\calD}{{\mathcal{D}}}

\newcommand{\calO}{{\mathcal{O}}}
\newcommand{\calP}{{\mathcal{P}}}

\newcommand{\calS}{{\mathcal{S}}}

\newcommand{\calW}{{\mathcal{W}}}

\SetKw{Break}{break}

\title{Tokenization: Hardness and Bicriteria Approximation Algorithms}
\author{
  Atieh Bastan\thanks{Atieh and Xiaoling are equal first authors.}\\
  Iowa State University.
   \texttt{atie@iastate.edu}
  \and
  Xiaoling He\footnotemark[1]\\
  University of Nebraska--Lincoln.
  \texttt{xhe16@huskers.unl.edu}
  \and
  A. Pavan\\
  Iowa State University.
  \texttt{pavan@iastate.edu}
  \and
  N. V. Vinodchandran\\
  University of Nebraska--Lincoln.
  \texttt{vinod@unl.edu}
}
\date{}
\begin{document}
\maketitle
\input{sections/abstract}

\newpage
\tableofcontents
\clearpage
\input{sections/introduction}

\input{sections/preliminaries}
\input{sections/binary-tokenization}
\paragraph{AI Disclosure.} 
The authors used ChatGPT Astra to explore proof strategies, generate ideas, and refine arguments. The authors have independently verified all results and proofs and take full responsibility for the correctness and content of this paper.

\clearpage
\phantomsection
\addcontentsline{toc}{section}{References}
\bibliographystyle{plainnat}
\input{sections/references}
\appendix
\input{sections/appendix}
\input{sections/index}
\end{document}

%% file: sections/abstract.tex
\begin{abstract}
We study \emph{Binary Tokenization}: given a corpus of binary strings
over $\{0,1\}$ and a vocabulary budget $K$, select at most $K$ substrings (called vocabulary) to minimize
the total number of tokens needed to represent the corpus. The
primitive symbols $0$ and $1$ are always available, and each document
must be segmented separately. Letting $n$ denote the total corpus length
and $M$ for the maximum document length, we prove that, for every
fixed $\varepsilon\in(0,1)$, Binary Tokenization problem does not have  polynomial-time
$n^{1-\varepsilon}$-approximation algorithms unless
$\mathsf{P}=\mathsf{NP}$.
To complement the hardness result, we give a polynomial-time
$(\gamma,\gamma)$-bicriteria algorithm, with $\gamma = \sqrt{M}\log n$:
The algorithm outputs a vocabulary of size at most $\gamma K$ with number of tokens at most $\gamma \OPT_K$, where $\OPT_K$ is the optimal number of tokens under the original budget $K$. 
We also study a storage-budget variant, in which the sum of the
selected substrings' lengths is bounded by $B$. For $2\le B<n$, we obtain an $(O(\log n\log B),O(\log n\log B))$-bicriteria approximation algorithm.

\end{abstract}

%% file: sections/introduction.tex
\section{Introduction}
\label{sec:introduction}
Large language models typically process text as sequences of {\em tokens} drawn from a fixed vocabulary. The process of segmenting text into units, called {\em tokenization}, determines the representation on which the model operates. Tokenization is a fundamental determinant of LLM's computational efficiency. Representing the same text with fewer tokens can reduce training and inference costs and allow more content to fit within a fixed context window; empirical studies show that tokenization can substantially affect these benefits as well as downstream performance. (\cite{dagan2024getting}; \cite{Ali2024}) Achieving shorter sequences, however, requires choosing an effective vocabulary: adding longer or more specialized tokens can improve compression, but enlarges the model’s embedding and output layers, increasing their memory and computational costs. (\cite{dagan2024getting}) This tradeoff motivates a basic optimization question underlying tokenization: given a limited vocabulary budget, which substrings should be selected to minimize the number of tokens needed to represent a corpus? Understanding the computational limits of this problem provides a theoretical foundation for an important efficiency objective in large language modeling.

We consider the optimization problem underlying this tradeoff. Given a corpus and a vocabulary budget, the goal is to select a shared set of substrings that minimizes the total number of tokens in the corpus. Each document must be partitioned into selected tokens, with individual characters always available. Although there are efficient algorithms for finding an optimal segmentation for a fixed vocabulary, choosing the vocabulary requires coordinating decisions across the entire corpus. A useful token may occur in many documents, and overlapping occurrences create competition among otherwise promising choices. Consequently, frequency alone does not capture the value of a vocabulary entry. 

The computational complexity  of tokenization has recently attracted
increasing attention. Whittington, Bachmann, and
Pimentel \cite{whittington2025npcomplete} proved that the decision versions of both
direct tokenization, which selects a vocabulary, and bottom-up
tokenization, which selects a sequence of pair merges, are NP-complete.
Kastreva et al.\cite{kastreva2026bounded} strengthened these results by
establishing APX-hardness even over a binary alphabet, ruling out
polynomial-time approximation schemes unless $\mathsf{P}=\mathsf{NP}$;
they also showed that direct tokenization remains NP-complete over a
unary alphabet. On the algorithmic side, Lim et al.~\cite{lim2025partition}
formulated vocabulary selection as a partition-cover problem and
introduced GreedTok, while Tempus et al.~\cite{tempus2026convex} developed
ConvexTok using convex relaxations that provide instance-specific
certificates of solution quality. Approximation guarantees have also
been established for compression savings: Kozma and
Voderholzer~\cite{kozma2026bpe} proved that BPE achieves at least one-third
of the optimal pair-encoding utility. These savings guarantees do not imply
corresponding multiplicative approximations for minimizing the
remaining token count.

Our main results establish polynomial-factor \(n^{1-\varepsilon}\) hardness of approximation for this problem for any $1> \varepsilon > 0$, even over a binary alphabet, where \(n\) denotes the total corpus length. Earlier work established only constant-factor inapproximability, ruling
out polynomial-time approximation ratios below $c=1.000002$ unless
$\mathsf{P}=\mathsf{NP}$~\cite{kastreva2026bounded}. 
Motivated by this difficulty, we develop bicriteria approximation algorithms that allow a controlled increase in vocabulary size while bounding the token count relative to the optimum under the original budget.
In particular, for a binary corpus of total length $n$ and maximum document length $M$,
we a polynomial-time $(\gamma,\gamma)$-bicriteria algorithms,
with 
$\gamma=O(\sqrt{M}\log n)$, where $K$ is the vocabulary
budget. Each algorithm selects at most $\gamma K$ vocabulary entries
and produces at most $\gamma\operatorname{OPT}_K$ tokens, where
$\operatorname{OPT}_K$ is the minimum token count under the original
budget $K$.

We also consider a storage-budget model, which requires
$\sum_{w\in V}|w|\le B$ instead of bounding the number of vocabulary
entries. For $2\le B<n$, we obtain a polynomial-time $(O(\log n\log B),O(\log n\log B))$-bicriteria approximation
for storage and token count, measured against the optimum with
storage budget $B$. 

We would like to note that it is easy to get a $n/\log n$-approximation algorithm: order the tockens by their frequency and take the top $K$ of them where $K$ is the budget on the cardinality of the vocabulary.

%% file: sections/preliminaries.tex
\section{Preliminaries and model definitions}
\label{sec:preliminaries}
\subsection{Binary Tokenization}
Let $s\in \Sigma^+$ be a nonempty string of characters $c$ from a binary alphabet $\Sigma$, we call it a document. A collection of documents form a corpus $\calD=\{s_i\}_{i=1}^d$ that is used to train language models. However, this corpus is considered raw and as a preprocessing step for a language model, we segment it before feeding it for training. During segmentation, a corpus is transformed into sequence of tokens suitable as an input. Let $\calW(\calD)=\{w:|w|\ge2,\ w\text{ occurs within some }s_i\}$ be a set of all possible tokens given $\calD$, $n=\sum_{i=1}^d|s_i|$ be the total length of the corpus and $M=\max_i|s_i|$ be the length of the longest string. 

For fixed $V\subseteq \calW$, let
\[
 \Tok(s,V)=\min\{t:s=w_1\cdots w_t,\ w_j\in V\cup\Sigma\},
 \qquad T(\calD,V)=\sum_i\Tok(s_i,V).
\]



\begin{definition}[Tokenization (Tok)]\label{def:binary}
Given $\calD$ and an integer $K\ge0$, minimize
\begin{equation}\label{eq:problem}
 \OPT_K(\calD)=\min_{V\subseteq\calW(\calD),\ |V|\le K}T(\calD,V).
\end{equation}
\end{definition}
In the literature, this is also called Binary Direct Tokenization.


\begin{definition}[Binary Tokenization with a storage budget]\label{def:storage}
Let $c(V)=\sum_{w\in V}|w|$. Given $\calD$ and an integer $B\ge0$, minimize
\[
 \OPT^{\mathrm{st}}_B(\calD)=
 \min_{V\subseteq\calW(\calD),\ c(V)\le B}T(\calD,V).
\]
\end{definition}
The storage cost charges each distinct spelling once, without shared
representation. It excludes the free primitive bits. The alphabet is
binary in both definitions; the occurrence-cover arguments themselves
also work over general finite alphabets.

An $(\alpha,\beta)$ bicriteria algorithm returns at most $\alpha K$ entries
and at most $\beta\OPT_K$ tokens for cardinality, or storage at most
$\alpha B$ and at most $\beta\OPT^{\mathrm{st}}_B$ tokens for storage.
The comparison optimum always obeys the original budget. A standard
approximation has $\alpha=1$. Let $H_n=\sum_{j=1}^n1/j\le1+\ln n$.
We use $\epsilon$ for the empty word and $\varepsilon$ for an approximation
exponent.

With $K=0$ or $B<2$, primitives are exact. In the cardinality model, one
string is trivial once $K\ge1$; a corpus is trivial when all its distinct
nonprimitive documents fit. In storage, $B\ge n$ permits buying all such
documents. The interesting storage range is therefore $2\le B<n$.
All complexity bounds use explicitly written corpus length, not succinct
binary-encoded document multiplicities.

\subsection{Relevant Definitions For Hardness Proof}
In the hardness proof, we will use the Vertex Cover decision problem as the source of the hardness, then use the Word Factorization as an intermediate step to establish a hardness factor for the Binary Tokenization optimization problem. The relevant definitions are given below.

\begin{definition}[Vertex Cover (\hypertarget{VC}{VC})]

\addcontentsline{toc}{subsubsection}
  {Definition~\ref{VC}: Vertex Cover}
Let $G=(Q,E)$ be a simple undirected connected graph where $Q$ is the set of vertices and $E$ is the set of edges. Denote the size of the vertices to be $w=|Q|$ and the size of the edges to be $m=|E|$. Let $\tau(G)$ denote the minimum size of a vertex cover. For a budget $\Phi$, The vertex cover decision problem outputs

\begin{itemize}
    \item \textbf{YES:} if $\tau(G)\leq \Phi$.
    \item \textbf{NO:} if $\tau(G)>\Phi$.
\end{itemize} 
\end{definition}

We will define the word factorization problem next. In this problem, one seeks to minimize a property of the given corpus called "rank". The rank counts the minimum number of generators needed to generate the corpus. It is analagous to the concepts of the dimension counting the size of a basis in linear algebra.

\begin{definition}[Generating set]
Let $A$ be an alphabet and let $X \subseteq A^*$ be a set of words.
A set $Y \subseteq A^*$ is a \emph{generating set} for $X$ if
every word in $X$ can be written as a finite concatenation of words
from $Y$. Equivalently,
\[
X \subseteq Y^*,
\]
where $Y^*$ denotes the set of all finite concatenations of elements
of $Y$, including the empty word.
\end{definition}

\begin{definition}[Generating Rank]
\label{rk}
Let $X\subseteq A^*$ be a finite set of nonempty words
over an alphabet $A$. The rank of $X$ is the size of its smallest generating set, i.e.
\[
\operatorname{rk}(X)
=
\min\{|Y| : Y\subseteq A^*,\; X\subseteq Y^*\}.
\]
It is important to note that singleton letters in $Y$ count toward $|Y|$.
\end{definition}

\begin{definition}[Word Factorization (FAC)]
\label{FAC}
\addcontentsline{toc}{subsubsection}
  {Definition~\ref{FAC}: Word Factorization}
Given a finite set of nonempty words $X \subseteq A^+$ over
an alphabet $A$ and an integer $B \ge 0$. The Word Factorization problem outputs
\begin{itemize}
    \item \textbf{YES:} if $rk(X)\leq B$.
    \item \textbf{NO:} if $rk(X)>B$.
\end{itemize}
\end{definition}

%% file: sections/binary-tokenization.tex
\section{Binary Tokenization}
\label{sec:binary}
\input{sections/binary-hardness}\label{sec:hardness}
\input{sections/binary-bicriteria-reorganized}

%% file: sections/binary-hardness.tex
\subsection{Hardness of approximation}

We will use the NP-hardness of the Vertex Cover decision problem to produce a hardness of approximation gap for the Binary Tokenization optimization problem. Step one is to first reduce VC to FAC and prove a relationship between the objective values of the two problems. This reduction is adapted from the paper "Elementariness of a finite set of words is co-NP-complete"~\cite{neraud1990elementariness}. Step two is, through a binary encoding function, transform the FAC instance into a Binary Tokenization instance and prove a relationship between the objective values of these two instances. Lastly, we would combine the two relationships between objective values to obtain the hardness factor of the Binary Tokenization problem.

\begin{reduction}[VC to FAC]
\label{red:vc-fac}
\addcontentsline{toc}{subsubsection}
  {Reduction~\ref{red:vc-fac}: Vertex Cover to Word Factorization}
Consider a \hyperlink{VC}{vertex cover} instance with graph $G=(Q,E)$ and budget $\Phi$. Assign distinct singleton letters to each vertex of $Q$. For each edge $e=(u_e,v_e)$, also assign 3 distinct edge-specific singleton letters $c_e,\lambda_e,\rho_e$. Let alphabet $A$ be the set consist of all these singleton letters. Note $|A|=w+3m$.

For each edge $e=(u_e,v_e)$ with edge specific singletons $c_e,\lambda_e,\rho_e$, we make a set of words 
$$X_e=\{u_e c_e v_e,u_e c_e \rho_e, \lambda_e c_e v_e, \lambda_e c_e \rho_e\}$$
Let our total set of words be 
$$X_G = \bigcup_{e\in E}X_e$$
Let $B=3m+\Phi$. Let $N$ denote the length of $X$, i.e. the sum of the lengths of all the words in $X$. Here, \hypertarget{N}{$N=12m$}. 

Note \hypertarget{A<N}{$|A|\leq N$}because $w\leq 2m$ because each vertex is adjacent to at least one edge due to connectedness.
\end{reduction} 

The following lemma provides a relation between $rk(X_G)$ and $\tau(G)$.

\begin{lemma}[$rk(X_G), \tau(G)$ identity]
\label{lem:source-rank}
The source corpus satisfies
\begin{align}
    \operatorname{rk}(X_G)=3m+\tau(G).
\end{align}

\end{lemma}

The proof of Lemma~\ref{lem:source-rank} is given in Appendix~\ref{pf:source-rank}.

The above lemma bridges the \textbf{YES} and \textbf{NO} instances of the two problems, we have
\begin{align}
&\tau(G)\leq \Phi \Longrightarrow rk(X_G)\leq B,\\
&\tau(G)> \Phi \Longrightarrow rk(X_G)> B.
\end{align}

The idea of the remaining of the proof is that, even after the encoding, the resulting Binary Tokenization problem is still in essence the original Word Factorization problem because we will process each letter's encoding as its own chunk, i.e. we avoid segmenting a letter's encoding. We will give each letter its unique binary encoding, and we combine (in a slightly sophisticated way) these binary encoded letters according to the original corpus $X_G$ to form the binary encoded corpus. We want the Tokenization instance to segment the binary encoded corpus corresponding to the original letters, i.e. we want a segmentation where the token boundaries align with the boundaries of the binary encodings of the letters. To ensure the alignment of the boundaries, what we do is to repeat each binary encoded letter a large enough number of times. One of these repeats will not have misaligned boundaries, then we tokenize all other repeats the same way as the "good" repeat. Since the boundaries align, no letter has its binary encoding cut into pieces, then we can factorize the original letter corpus according to binary tokenization of the binary encoded corpus. 

\begin{Transformation}[FAC instance to Tok instance]
\label{Transformation1}
\addcontentsline{toc}{subsubsection}
  {Transformation~\ref{Transformation1}: Word Factorization instance to Tokenization instance}
Let the FAC instance be the one defined in \hyperref[red:vc-fac]{Reduction~\ref*{red:vc-fac}}. We will now encode it into a binary corpus. Let $\varepsilon$ be the constant defined in \hyperref[thm:main]{Theorem~\ref*{thm:main}}. Define \hypertarget{param}{$q:=\lceil 4/\varepsilon\rceil$, $p:=N+2q+10$, $R:=p^q$.} We define $R$ here because we eventually want to repeat each letter in the FAC problem $R$ times in the final encoding. 

The "building blocks" to the final binary encoding are the elements $(a,j)\in A\times \{1,...,R\}$. As you can see, we are repeating each element $a\in A$ a total of $R$ times. For an element $(a,j)$, we call $j$ its "phase". For every $(a,j)\in A\times \{1,...,R\}$, we want to assign it a unique binary string. Since there are $|A|\cdot R$ number of such elements, we will need \hypertarget{b}{$b=\max(1,\lceil\log_2(|A|\cdot R)\rceil)$} bits in the binary string. Choose any \hypertarget{inj}{injective function} \hypertarget{z}{$z(a,j):A\times \{1,...,R\}\longrightarrow \{0,1\}^b$} to do the assignment. Such a function exists because, for example, one can naively enumerate the elements and assign to each element their enumeration. 

For an output string $z=z_1...z_b$ of the function $z$ chosen above, we add delimiters and alternating internal padding to it as follow 
$$C(z):=110z_10z_20...0z_b0$$
Finally, we will define the unique "binary block" associated to element $(a,j)$ to be
$$\hypertarget{bb}{B_{a,j}:=C(z(a,j))}$$
Note each binary block has length \hypertarget{l}{$l=2b+3$}.

Now we can define the binary encoding function $h_R:A^+\longrightarrow \{0,1\}^+$. For a singleton letter $a\in A$, we define the binary encoding to be the concatenation of all $R$ "repeats" of the $B_{a,j}$ blocks.
$$h_R(a):=B_{a,1}B_{a,2}...B_{a,R}$$
Note each encoding of a singleton letter has length $Rl$. Now for a word $x\in A^+, x=a_1...a_r$, we just define its binary encoding to be the concatenation of the binary encoding of each of its singleton letters.
$$\hypertarget{hG}{h_R(x) := h_R(a_1)...h_R(a_r)=B_{a_1,1}...B_{a_1,R}B_{a_2,1}...B_{a_2,R}...B_{a_r,1}...B_{a_r,R}}$$

Finally, we can now define the corpus of the tokenization problem.
$$D_R := \{ h_R(x) : x\in X_G \}$$
This corpus has $d = 4m$ distinct documents of length $3Rl$ and total length  
\begin{align}
    n=NRl=12mRl \label{ngl}
\end{align}.

Let the alphabet be $\Sigma = \{0,1\}$. Let $K=B=3m+\Phi$.

\end{Transformation}

Building off of \hyperref[lem:source-rank]{Lemma~\ref*{lem:source-rank}}, the following proposition relates $\tau(G)$ and $OPT_K(D_R)$. Note that at this step, the choice of $R$ is not yet important.

\begin{proposition}
\label{prop:gap}
For every integer $R \ge 1$, \hyperref[Transformation1]{Transformation~\ref*{Transformation1}} satisfies

\begin{align}
rk(X_G) \le B
&\implies \operatorname{OPT}_K(D_R) \le 8m,\\
rk(X_G) > B
&\implies \operatorname{OPT}_K(D_R) \ge R+4m.
\end{align} 
\end{proposition}

The proof is given in Appendix~\ref{pf:gap}.

Applying \hyperref[lem:source-rank]{Lemma~\ref*{lem:source-rank}} with \hyperref[prop:gap]{Proposition~\ref*{prop:gap}}, we get 

\begin{align}
&\textbf{YES: }\tau(G) \le \Phi
\implies \operatorname{OPT}_K(D_R) \le 8m,\label{yes}\\
&\textbf{NO: }\tau(G) > \Phi
\implies \operatorname{OPT}_K(D_R) \ge R+4m.\label{no}
\end{align} 

At last, we obtain the result for our main theorem.

\label{sec:main}
\begin{theorem}[hardness of approximation]
\label{thm:main}
\addcontentsline{toc}{subsubsection}
  {Theorem~\ref{thm:main}: hardness of approximation of Tok}
For every fixed $\varepsilon \in (0,1)$, unless
$\mathrm{P}=\mathrm{NP}$, there is no deterministic
polynomial-time algorithm that returns, on every input,
a vocabulary $V$ at most size $K$ satisfying
\[
T(D,V) \le n^{1-\varepsilon}\operatorname{OPT}_K(D),
\]
where $n$ is the total corpus length.
This holds even when all documents are distinct binary
strings of the same length.
\end{theorem}

\begin{proof}
\label{main_proof}
\addcontentsline{toc}{subsubsection}
  {proof~\ref{main_proof}: Proof of Theorem~\ref{thm:main}}
    Assume for the sake of contradiction, such approximation algorithm exists that approximates $T(D,V)$ to within $n^{1-\varepsilon}\operatorname{OPT}_K(D)$ with a vocabulary $V$ at most size $K$. Take a nontrivial Vertex Cover instance and construct $D_R$ as above. Run the algorithm at budget $K=3m+\Phi$. Run the polynomial PathPiece~\cite{goldman2024unpacking} algorithm, which outputs the optimal segmentation given a vocabulary, on the output vocabulary to obtain $T(D_R,V)$.

    Before we do further computation, let's examine some relationships between parameters. Recall the definition of parameters \hyperlink{param}{ $q,p,R$}, \hyperlink{b}{$b$} and \hyperlink{N}{$N$} and the fact that \hyperlink{A<N}{$|A|\leq N$}. Due to the definition of $p$, we have $|A|\leq N\leq p$ and $p\geq 3$, so we can obtain
    \begin{align}
        b\leq (q+1)\log_2p+1\leq (q+1)p+1\\
        \hyperlink{l}{l=2b+3}\leq (2q+2)p+5\leq (2q+4)p<p^2.\label{l}
    \end{align} 

    Combining equation~\eqref{ngl} and equation~\eqref{l}, we get
    
    \begin{align}
        n=NRl\leq Rp^3=p^{q+3}\label{nbound}
    \end{align} 
    In a YES instance, we would have
    
    \begin{align}
        T(D_R,V) &\leq n^{1-\varepsilon} \operatorname{OPT}_K(D_R)\\
        &\leq n^{1-\varepsilon} 8m\\
        &\leq (Rp^3)^{1-\varepsilon} p\\
        &=R^{1-\varepsilon}p^{4-3\varepsilon}<R
    \end{align} 
    The second inequality is due to equation~\eqref{yes}, the third inequality is due to equation~\eqref{nbound} and the fact that $8m\leq N\leq p$. To justify the last inequality, $$R^\varepsilon=p^{q\varepsilon}\geq p^4>p^{4-3\varepsilon}$$.

    In a NO instance, every feasible output satisfies $T\geq R+4m>R$ by equation~\eqref{no}.

    Note $T<R$ and $T>R$ partitions the possibilities of being in the YES or NO case for the Vertex Cover decision problem. Therefore, testing the integer inequality $T<R$ decides Vertex Cover in polynomial time. We've reached a contradiction. Hence, we've proven Theorem~\ref{thm:main}.
\end{proof}

%% file: sections/binary-bicriteria-reorganized.tex
\subsection{Bicriteria approximation algorithms}
\label{sec:card-bicriteria}
In section \ref{sec:hardness}, we showed the hardness of approximating the tokenization problem and its NP-hardness. The following results provide a polynomial time bicriteria approximation algorithms that bounds the vocabulary size and the tokenization cost within guaranteed multiplicative factors of the optimal solution, covering nontrivial cardinality range $1\leq K \leq n$. We write $\operatorname{OPT}_K=\OPT_K(\calD)$ and use the corpus parameters
of Section~\ref{sec:preliminaries}. Also, whenever tokens are mentioned, they refer to $w\in\calW$.

An occurrence of a word $w\in\calW$ in a document $s\in\calD$
is an interval $(s,[i,j))$ such that
$w$ appears in positions $i,\ldots,j-1$ in $s$, and we say the interval covers $[i,j)$.
Two occurrences $(s_1,[i_1,j_1))$ and
$(s_2,[i_2,j_2))$ are disjoint if they lie in different
documents, or if their intervals do not overlap. 
Let $o_a=\sum_{b<a}|s_b|$ be the offset of document $s_a$. We map each occurrence to its interval in corpus coordinates by
\[
    f\bigl((s_a,[i,j))\bigr)=[o_a+i,o_a+j).
\]
Therefore, each occurrence is identified with its global interval and denoted by $I$.
For $w\in\calW$, define
\[\calO_w=\left\{f\bigl((s_a,[i,j))\bigr)\;\middle|\;\begin{array}{l}1\le a\le d,\quad 1\le i<j\le |s_a|+1,\\
s_a[i]s_a[i+1]\cdots s_a[j-1]=w
 \end{array}
\right\}.
\]
Thus, $\calO_w$ is the set of all occurrence intervals
of $w$ in the corpus.

A \emph{bundle} for $w$ is a nonempty subset $Q_w\subseteq \calO_w$ whose occurrences are pairwise disjoint. For a fixed parameter $\lambda>0$, the cost of a bundle $Q_w$ is:
\[c(Q_w)=\lambda + |Q_w|\]
That is, selecting a bundle incurs a fixed cost $\lambda$
and a unit cost for each occurrence it contains. Given a set $U\subseteq\{1,\dots,n\}$, \emph{gain} and \emph{density} of a bundle $Q_w$ is 
\[g_U(Q_w)=|U\cap \bigcup_{I\in Q_w} I|,\qquad \rho_U(Q_w)=\frac{g}{c(Q_w)}\]
We also allow a corpus position to be covered by a single character $c\in\Sigma$ and we call it a singleton, with cost, density and gain of $1$. 

A \emph{cover} $\calC=(\calC_\calW,\calC_\Sigma)$ consists of a list of bundles and a list of singletons such that their union covers every corpus position. The cost of a cover is the sum of the costs of its listed bundles and singletons. That is:
\[c(\calC)=\sum_{Q\in\calC_\calW}c(Q)+|\calC_\Sigma|.
\]

For each $w$, order its $m_w=|\mathcal O_w|$ occurrences
$I_1,\ldots,I_{m_w}$ by increasing global start coordinate, writing
$I_i=[a_i,b_i)$. For each interval $I_i$, define
\[
 p_i=\max\bigl(\{0\}\cup\{k<i:b_k\le a_i\}\bigr)
\]
to be the index of the last occurrence that doesn't overlap with $I_i$.
For a fixed $U\subseteq \{1,\dots,n\}$ and $w\in\calW$, let $g_i= |U\cap I_i|$. Then, define $F_w(i,j)$ to be the maximum total gain of exactly $j$ disjoint interval occurrences of $w$ among $I_1,\dots,I_i$. For every $0\leq j\leq i\leq m_w$:
\[
F_w(i,j)= \max\{\sum_{I_k\in Q}g_k| Q\subseteq \{I_1,\dots,I_i\}, |Q|=j, Q\text{ is pairwise disjoint}\}
\]
We compute this value using a two dimension dynamic program, initialized as:
\[
 F_w(i,0)=0,\qquad F_w(0,j)=-\infty\quad(j>0).
\]
Then, use
\[
 F_w(i,j)=\max\{F_w(i-1,j),\ g_i+F_w(p_i,j-1)\}.
\]

The first term excludes $I_i$ and selects $j$ disjoint
occurrences among the first $i-1$ occurrences.
The second term includes $I_i$ and selects the remaining
$j-1$ occurrences among the first $p_i$ occurrences.

We can get the best-density bundle for $w$ by:
\[ \max_{1\leq j\leq m_w} \frac{F_w(m_w,j)}{\lambda+j}\]
The algorithm~\ref{algo:FWB} finds the best density and by backtracking, identifies a bundle with this density.
\begin{algorithm}\label{algo:FWB}
\SetKwInOut{Output}{Output}
\SetKwInOut{Input}{Input}
\Input{Ordered occurrence set $\calO$, Uncovered positions for each occurrence $G$, Predecessor indices $P$, $\lambda>0$}
\Output{A maximum-density bundle $Q$ and its density,
        or $(\emptyset,0)$ if no bundle has positive gain}
$m \gets len(\calO), F\gets$ $m\times (m+1)$ table initialized at $-\infty$\\
\For{$i \gets 1 $\KwTo $m$}{
$F(i,0)\gets 0$\\
}
\For{$i\gets 1$ \KwTo $m$}{
\For{$j\gets 1$ \KwTo $i$}{
$F(i,j)\gets \max\{F(i-1,j), G[i]+F(P[i],j-1))\}$\\
}
}
$\rho^*\gets0$ and $j^*\gets0$\\
\For{$j\gets 1$ \KwTo $m$}{
$\rho\gets \frac{F(m,j)}{\lambda+j}$\\
\If{$\rho > \rho^*$}{
$\rho^*\gets \rho, j^*\gets j$
}
}
$Q\gets\emptyset$, $j\gets j^*$, $i\gets m$\\
\While{j > 0}{
\eIf{F(i,j) > F(i-1,j)}{
$Q.append(\calO[i]), i\gets P[i], j\gets j-1$\\
}{
$i\gets i-1$\\
}
}
\Output {$Q$, $\rho^*$}
\caption{Fixed Word Best Bundle} 
\end{algorithm} 

We, then, construct a greedy cover, starting with $U=\{1,\dots,n\}$, the set of uncovered corpus positions. At each iteration, we run $\operatorname{FixedWordBundle}$ exhaustively for all $w\in\calW$, with gain computed in relation to $U$, and select the max-density bundle. If the selected bundle's density exceeds $1$, we add this bundle to the cover. Otherwise, we add an uncovered position as a singleton. After each selection, we remove the newly covered positions from $U$ and repeat until $U=\emptyset$. Since each iteration at least covers one position, this procedure ends after at most $n$ iterations. The selected bundles and singletons form a cover $\calC=(\calC_\calW,\calC_\Sigma)$ of corpus. This process is described in algorithm~\ref{algo:Greedy_bundle}.

\begin{algorithm}\label{algo:Greedy_bundle}
\SetKwInOut{Output}{Output}
\SetKwInOut{Input}{Input}
\Input{The corpus $\calD$, corpus size $n$, sorted occurrence list for each candidate words $\calO$, Predecessor indices table $P$, $\lambda>0$}
$U\gets \{1,\dots,n\}, \calC_\calW\gets\emptyset, \calC_\Sigma\gets\emptyset, R\gets0 , V\gets\emptyset$\\
\While{ $U \ne \emptyset$}{
$\rho^*\gets 1, Q^*\gets\emptyset,w^*=\bot$\\
\ForEach{$w\in\calW$}{
\For{$i\gets 0$ \KwTo len($\calO[w]$)}{
$G[w][i]\gets |U\cap \calO[w][i]|$\\}
$Q, \rho\gets \operatorname{FixedWordBestBundle}(\calO[w],G[w],P[w],\lambda)$\\
\If{$\rho >\rho^*$}{
$\rho^*\gets \rho, Q^*\gets Q, w^*\gets w$
}
}
\eIf{$w^*==\bot$}{
$x\gets \min U$\\
$\calC_\Sigma.append(x), J \gets \{x\}, R\gets R+1$\\
}{
$J\gets U\cap\bigl(\bigcup_{I\in Q^*}I\bigr)$\\
$\calC_\calW.append(Q^*),  R\gets R + len(Q^*), V \gets V \cup \{w^*\}$
}
$U \gets U \setminus J$
}
\Output{$\calC_\calW, \calC_\Sigma, R,V$}
\caption{Greedy Bundle Cover} 
\end{algorithm} 

Although occurrences within one bundle are pairwise disjoint, occurrences from different bundles may overlap. Thus, the greedy cover $\calC$ from algorithm~\ref{algo:Greedy_bundle} may not be a valid segmentation. We next show how to convert this cover into a valid segmentation and give an upper bound on its vocabulary size and token counts.

First, we define a \emph{chunk construction}\label{cons:chunks}. Fix an integer $1\leq q\leq M$. For a word $w\in\calW$ of length $\ell$, set $r_w=\lceil \frac{\ell}{q}\rceil$ and split $w$ from left to right into chunks of length $r_w$, with potentially shorter last chunk. Let $\operatorname{Chunk_q}(w)$ be the set of distinct chunks of length at least two of $w$. 

Let $V_g$ be the set of distinct words represented by the bundles in the greedy cover $\calC$. We construct the output vocabulary as
\[
    V=\bigcup_{w\in V_g}\operatorname{Chunk}_q(w).
\]
We can see that $V\subseteq\calW$.

\begin{algorithm}\label{Final_algo}
\SetKwInOut{Output}{Output}
\SetKwInOut{Input}{Input}
\Input{corpus $\calD$, list of candidate words $\calW$,$q$, vocabulary budget $K,\eta > 0$}
\Output{Vocabulary $V$, a segmentation $\calP$ and its token count $T(\calD,V)$}
$n\gets \sum_{s\in\calD} len(s)$\\
\ForEach{$w\in\calW$}{
$\calO[w]\gets$ occurrences of $w$ in $\calD$\\
$\calO[w].sorted$ based on starting point\\
\For{$i\gets 1$ \KwTo $len(\calO[w])$}{
$P[w][i]\gets \max\bigl(\{0\}\cup \{k<i:\text{the endpoint of $\calO[w][k]\leq$ starting of }\calO[w][i]\}\bigr)$\\
}
}
\For{$t \gets 1$ \KwTo $n$}{
$\lambda\gets \frac{t}{K\eta}$\\
$\calC_\calW, \calC_\Sigma, R, V_g \gets \operatorname{GreedyBundleCover}(\calD, n, \calO, P, \lambda)$\\
$A \gets len(V_g)$\\
\If {$\lambda A+R\le (1+\ln{n}) t(1+\frac{1}{\eta})$}{
\Break
}
}
$V\gets\emptyset$\\
\ForEach{$w \in V_g$}{
$V\gets V\cup \operatorname{Chunk(q,w)}$\\
}
$\calP\gets \operatorname{MinTokenSegmentation}(\calD,V\cup\Sigma), t\gets len(\calP)$\\
\Output {$V, \calP, t$}
\caption{Final Algorithm} 
\end{algorithm} 

\subsection{Analysis}

\begin{theorem}[Substring-chunk bicriteria guarantee]\label{thm:ucb-chunks}
Let $\calD$ be a corpus with $M=\max_{s_i\in \calD}|s_i|$ and $\calW$ be a set of all possible tokens. For any cardinality budget $K \geq 1$, max string size $M\geq 1$ and any integer $1\leq q \leq M$ and a rational $\eta>0$, Algorithm~\ref{Final_algo} runs in deterministic
polynomial time and returns a vocabulary $V\subseteq \calW$ and valid segmentation of $\calD$ satisfying
\[
|V|\leq (1+\eta)(1+\ln{n}) q K \qquad \text{and} \qquad T(\calD, V)\leq (1+\frac{1}{\eta})(1+\ln{n}) h \OPT_K\\
\]
where $h= q-1+\lceil \frac{M}{q}\rceil$.
\end{theorem}

\begin{definition}[Segmentation cover]\label{comp_cover}
Let $V$ be a vocabulary and fix a segmentation $\calP$ of $\calD$ over $V\cup\Sigma$. For each $w\in V$, let $Q_w$ contain the occurrences of $w$ in $\calP$. These occurrences are pairwise disjoint, so every nonempty $Q_w$ is a  bundle.

Let
\[
\calB=\{Q_w:w\in V,\ Q_w\ne\emptyset\},
\]
and let $\calS$ be the set of positions covered by
singletons in $\calP$.
The \emph{segmentation cover} induced by $\calP$ is
the pair $(\calB,\calS)$.
We call $\calB$ its bundle part and $\calS$
its singleton part.
\end{definition}
\begin{lemma}[segmentation  cost]\label{lem:ucb-segmentation}
Let $V$ be a vocabulary with $|V|\leq K$ and let $\calP$ be the segmentation cover induced by segmentation of $\calD$ over $V\cup\Sigma$ using $t=T(\calD, V)$ tokens. The cost of $\calP$ is at most $\lambda K + t$. 
\end{lemma}
\begin{proof}
Let $\calP=(\calB,\calS)$ where $\calB$ is the bundle part and $\calS$ is the singleton part of the segmentation cover. Define $V'=\{w| w\in V, Q_w\in \calB\}$ with $k=|V'|\leq |V|=K$. Since each token in the segmentation either contributes one occurrence to a bundle or a singleton-covered position to $\calS$, we get:
\[
\sum_{w\in V'}|Q_w| + |\calS| = t.
\]
Consequently, the cost of the cover is:
\begin{align*}
c(\calP)
&=\sum_{w\in V'}c(Q_w)+|\calS|\\
&=\lambda k+\sum_{w\in V'}|Q_w|+|\calS|\\
&=\lambda k+t\\
&\le\lambda K+t.
\end{align*}
\end{proof}

\begin{lemma}[Greedy cost]\label{lem:ucb-greedy}
Fix a positive rational $\lambda$. Let $V^*$ be an optimal budget vocabulary with $|V|\leq K$ and let $\OPT_K=T(\calD,V^*)$. Let $C$ be the cost of the induced segmentation cover. The greedy algorithm returns a cover $\calC=\calC_\calW\cup\calC_\Sigma$ with a total $R=\sum_{(w,Q)\in\calC_{\calW}}|Q|+|\calC_\Sigma|$ selected occurrences with a cost of at most  $ (1+\ln{n}) (\lambda K + \OPT_K)$.
\end{lemma}
\begin{proof}
Consider a greedy iteration with $r>0$ uncovered positions and the segmentation cover induced by $V^*$ has cost $C$. Since this cover comes from a segmentation, its bundles and singletons parts cover each of the remaining positions exactly once. Therefore, the sum of their newly covered position adds up to $r$. By averaging argument, at least one  has cost per newly covered position ratio of at most $\frac{C}{r}$. The greedy algorithm maximizes density, the reciprocal
of this ratio, so its selected bundle or singleton
also has cost per newly covered position at most $\frac{C}{r}$.

Fix iteration $i$, let $r_i$ be the number of uncovered positions before the selection and let $z_i$ be the number of newly covered positions, and $C_i$ the  cost of the selection. Then
\[\frac{C_i}{z_i}\leq \frac{C}{r_i}\implies C_i\leq C\frac{z_i}{r_i}\leq C\sum_{j=r_i-z_i+1}^{r_i} \frac{1}{j}\]
Each iteration covers at least one new position, so the algorithm terminates after $L\leq n$ iterations. Since $r_1=n$ and for all $i$, $r_{i+1}=r_i-z_i$ with $r_{L+1}=0$. Hence, 
\[\sum_{i=1}^L C_i \leq C\sum_{i=1}^n \frac{1}{i}\leq C(1+\ln{n})\]
By definition of the covering costs and $R$ and lemma~\ref{lem:ucb-segmentation}, we get
\begin{align}
 c(\calC)
 &=\lambda|\calC_{\calW}| + R\notag\\
    &= \sum_{i=1}^{L} C_i \notag\\
    &\leq (1+\ln n)C \notag\\
    &\leq (1+\ln n)(\lambda K+\OPT_K).
    \label{eq:ucb-cover}
\end{align}
\end{proof}

For a string $w\in \Sigma^*$, define
    \[\operatorname{Sub}(w)=\{u\in\Sigma^{>1}|\exists x,y\in\Sigma^* : w=xuy\}\quad\text{ and}  \quad \operatorname{Suf}(w)=\{u\in\Sigma^+| \exists x\in \Sigma^*: w=xu\}\]
    That is, $\operatorname{Sub}(w)$ is the set of all substrings of $w$ of length at least $2$, and $\operatorname{Suf}(w)$ is the set of all nonempty suffixes of $w$.

\begin{lemma}[A vocabulary of suffixes]\label{lem:ucb-suffix}
Fix $q\in\{2,\dots,M\}$. For every word $w\in\calW$ of length $l\leq M$, there exists a vocabulary $V_w\subseteq \operatorname{Sub}(w)$ such that $|V_w|\leq q$ and every $y\in \operatorname{Suf}(w)$ can be segmented into at most $h=q-1+\lceil\frac{M}{q}\rceil$ tokens from $V_w\cup\Sigma$.
\end{lemma}
\begin{proof}
Let $V_w$ be the set of distinct chunks construction by chunk construction in~\ref{cons:chunks}. Then $V_w\subseteq\operatorname{Sub}(w)$, and $|V_w|\leq q$. Every chunk, therefore, is a token in $V_w\cup\Sigma$.

For every nonempty suffix $y\in Suf(w)$ consider two cases:

\textit{Case 1}: $y$ begins at the start of a chunk. Then, $y$ consist of at most $b\leq q$ chunks from $V_w$. Since $q\leq h$, this gives a segmentation using at most $h$ tokens.

\textit{Case 2}: $y$ begins strictly inside a chunk. The portion of $y$ within this chunk has length at most $r-1$, and can be segmented into at most $r-1$ singletons. The rest of $y$ consist of at most $q-1$ chunks, each a token in $V_w$. Thus, the total number of tokens adds up to at most:
\[
 (r-1)+(q-1)\le q-1+\lceil M/q\rceil=h.
\]
Therefore, every suffix of $w$ admits a segmentation into at most $h$ tokens from $V_w\cup\Sigma$.
\end{proof}

\begin{lemma}[Cover to partition]\label{lem:ucb-partition}
Fix $q\in\{2,\dots,M\}$. Let \(\calC=\calC_{\calW}\cup\calC_\Sigma\) be a cover of $\calD$, where $\calC_\calW$ and $\calC_\Sigma$ are its  bundle and the singleton parts, respectively. Let $R$ be its total number of the selected occurrence and $V_g$ the distinct vocabulary. For each $w\in V_g$, let $V_w$ be the vocabulary given by lemma~\ref{lem:ucb-suffix}, and define
\[V=\bigcup_{w\in V_g}V_w.
\]
Then, $\calC$ admits a valid segmentation over $V\cup\Sigma$ using at most $hR$ tokens. Moreover, $|V|\leq q|V_g|\leq q|\calC_\calW|$.
\end{lemma}
\begin{proof}
    Fix a document $s\in\calD$. Let $x$ be its first unprocessed position, initially $1$. Since $\calC$ covers all positions of the corpus, there exists a selected occurrence interval $I=[a,b)$ such that $a\le x<b$. There are two cases:

    \textit{Case 1}: $I$ belongs to a selected bundle $I\in Q_w$ for some $Q_w\in\calC_\calW$. Then, the substring $s_x\dots s_{b-1}$ is a nonempty suffix of $w$. By lemma~\ref{lem:ucb-suffix}, it can be segmented into at most $h$ tokens from $V_w\cup\Sigma$, and therefore, $V\cup\Sigma$.
    
    \textit{Case 2}: $I$ is a selected singleton. Then, the substring $s_x$ contributes exactly one token, which is at most $h$.
    
    Add the resulting tokens to the segmentation, set $x=b$, and continue this process. Each step strictly increases $x$, and the emitted substrings are consecutive and remain within $s$. Coverage ensures that the process continues until $x=|s|+1$, so these substrings partition $s$.
    
    We can also ensure that no occurrence interval is used twice in this process: after choosing $I=[a,b)$, any subsequent unprocessed position is at least $b$.

    Repeating this process for all documents, uses at most $R$ selected occurrence intervals, each contributing at most $h$ tokens, giving at most $hR$ tokens.

    Finally, each $V_w$ contains at most $q$ chunks/tokens, so
    \[|V|=\left|\bigcup_{Q_w\in\calC_{\calW}}V_w\right| \leq\sum_{Q_w\in\calC_{\calW}}|V_w|\leq q|\calC_{\calW}|.\]
\end{proof}
\begin{proof}[Proof of Theorem~\ref{thm:ucb-chunks}]
Let $\alpha_n= (1+\ln{n})$. The greedy cost analysis gives:
\[
 \lambda A+R\le \alpha_n(\lambda K+\OPT_K),
\]
Since $1\leq\OPT_K\leq n$, the guess $t=\OPT_K$ is among those considered by the algorithm. At this guess, $\lambda=\frac{\OPT_K}{K\eta}$, so
\[\lambda A+R \leq\alpha_n\OPT_K(1+1/\eta).\]
Thus, the acceptance condition holds at $t=\OPT_K$. Because the guesses are considered in increasing order, the first accepted guess $\hat{t}$ satisfies $\hat{t}\leq\OPT_K$.

Let $A=|V_g|$ and $R$ be the occurrence counts of the accepted cover. The acceptance condition gives
\[\lambda A+R\leq\alpha_n\hat{t}(1+\frac{1}{\eta}),\qquad\lambda=\frac{\hat{t}}{\eta K}.
\]
Therefore,
\begin{align*}
A&\leq\frac{\alpha_n\hat{t}(1+\frac{1}{\eta})}{\frac{\hat{t}}{\eta K}}=(1+\eta)\alpha_n K,\\
R&\leq\alpha_n\hat{t}(1+\frac{1}{\eta})\leq(1+\frac{1}{\eta})\alpha_n\OPT_K.
\end{align*}

By Lemma~\ref{lem:ucb-partition}, the chunk vocabulary $V$ satisfies $|V|\leq qA$ and admits a valid segmentation using at most $hR$ tokens. Consequently,
\[ |V|\leq(1+\eta)(1+\ln n)qK\]
and
\[T(\calD,V)\leq(1+\frac{1}{\eta})(1+\ln n)h\OPT_K.\]
Every word in $V$ is a substring of a selected word and hence a substring of a corpus document. Since these words have length at least two, $V\subseteq\calW$.
\end{proof}
\begin{remark}
    The algorithm~\ref{Final_algo} achieves $(\gamma_q,\gamma_q)$-bicriteria approximation by choosing $\eta= \frac{h}{q}$ and $q=\lceil \sqrt{M}\rceil$, making $\gamma_q= O(\sqrt{M}\log n)$.
\end{remark}
\paragraph{Runtime analysis of Algorithm \ref{Final_algo}}: There are at most $O(n^2)$ words in $\calW$. Enumerating words, creating their occurrence lists, and ordering them can be done in polynomial time. In each greedy iteration, for every word $w\in\calW$, it takes $O(n^2)$ to compute $F_w(m_w,j)$ and a total of $O(n^4)$ to find the best overall bundle across all words and all sizes. During each iteration, at least one previously uncovered position is covered with at most $n$ iterations, totaling to $O(n^6)$ overall. Any selected collection of bundle contains at $O(n^2)$ occurrences, making the chunk vocabulary construction and segmentation all be done in polynomial time. 

\begin{corollary}\label{cor:ucb-best}
For $1\le K\le n$ and $M\ge2$, we can achieve equal bicriteria factors
\[
 \min\left\{1+\frac12\sqrt{n/K},\quad
 \ln{n}\min_{1\le q\le M}\bigl(2q-1+\lceil M/q\rceil\bigr)\right\}.
\]
\end{corollary}

\subsection{Storage Budget}
The above approach inspired us to formulate the direct binary tokenization problem in term of total storage budget. This problem is defined in Definition~\ref{def:storage}. We show the following.

\begin{theorem}\label{thm:storage-bicriteria}
Let $\calD$ be the corpus with total length $n$. Let $c(V)=\sum_{w\in V}|w|$. For $2\le B<n$, set $D_B=\lceil\log_2 B\rceil$ and any positive rational $\eta$, there is a deterministic polynomial-time algorithm that returns a vocabulary and valid segmentation with
\[
 c(V)\le(1+\eta)\ln{(n)}D_B B,\qquad
 T(\calD,V)\le(1+1/\eta)\ln{n}D_B\OPT^{st}_B.
\]
In particular $\eta=1$ gives equal factors
$2\ln{n}\lceil\log_2 B\rceil=O(\log n\log B)$.
\end{theorem}
Proof of this theorem is in Appendix~\ref{pf:storage}.

%% file: sections/appendix.tex
\appendix

\section{Proof of Lemma~\ref{lem:source-rank}}
\label{pf:source-rank}

To prove Lemma~\ref{lem:source-rank}, we need to first prove another lemma about a property of the corpus $X_G$. 

\begin{lemma}[Min Generators]
\label{lem:mingen}
    Let $u, v, c, \lambda, \rho$ be five distinct letters, and consider
    $$X = \{ucv, uc\rho, \lambda cv, \lambda c\rho\}.$$
    If one or both of $u$ and $v$ are available, then at least 3 extra generators are needed to generate $X$. 
    
    If neither $u$ nor $v$ are available, at least 4 generators are needed to generate $X$.
\end{lemma}

\begin{proof}
    To simplify the argument, suppose we have both $u$ and $v$ available (note restricting to only one of $u,v$ being available would only increase the number of generators needed). Assume for the sake of contradiction, we only need 2 extra generators. We'll discuss 2 cases: $\lambda c \rho$ is not a factor and $\lambda c \rho$ is a factor.

    In the first case, since $\lambda c \rho$ is not a factor, to factorize it, it needs at least 2 factors. Since we assume we only need 2 extra factors, we have to factorize $\lambda c \rho$ into exactly 2 factors, i.e. either $\{\lambda c, \rho\}$ or $\{\lambda, c \rho\}$. Together with the current factors $u$ and $v$ from assumption, none of these 2 sets of factors can factorize the word $ucv$. We've reached a contradiction.

    In the second case, since $\lambda c \rho$ is a factor, we can choose 1 extra factor. To factorize the word $ucv$, we need another generator. This generator must contain $c$ and be a factor of $ucv$. The possibilities are $c,uc, cv, ucv$. Together with the current factors $u,v,\lambda c \rho$ from assumption, none of these factors can factorize $\lambda cv$. We've reached a contradiction.

    Hence, even with both $u$ and $v$ available, we need at least 3 extra generators.

    Now suppose neither $u$ nor $v$ are available. Assume for the sake of contradiction that at most 3 generators suffice. Note at least 1 generator must start with $u$ and another start with $\lambda$. Also at least 1 generator must end with $v$ and another end with $\rho$. Since we're only allowed 3 generators, these generators must overlap. A factor that starts in $\{u, \lambda\}$ and ends in $\{v, \rho\}$ must be one of the full length-three words. 

    Remove such a full-word generator and the unique source word equal to it. It cannot occur in any of the three remaining source words, since those words are distinct and have the same length. Thus at most two remaining generators must generate those three words. The three words still have both initial letters $u,\lambda$ and both final letters $v,\rho$. So by the logic in the last paragraph, we must have at least 1 generator starting with $u$ and another starting with $\lambda$, and at least 1 generator ending with $v$ and another ending with $\rho$. Since we're only allowed 2 generators here, we see that both generators are full length-three words. But two full length-three words cannot generate three full length-three words. We've reached a contradiction. 

    Hence, with neither $u$ nor $v$ available, we need at least 4 generators.
\end{proof}

\textbf{Lemma~\ref{lem:source-rank}.}
\textit{The source corpus satisfies}

\begin{align}
    \operatorname{rk}(X_G)=3m+\tau(G).
\end{align} 


\begin{proof}
    We'll prove something stronger: any vertex cover $U$ yields a generating set of size $3m+|U|$ and a factorization of every source word into exactly two factors.

    To prove 
    $\operatorname{rk}(X_G)\leq 3m+|U|$, first select all the singleton vertex in $U$. Do the following to factorize $X_e\subset X_G$ for any edge $e=(u_e,v_e)$. If $u_e\in U$, select in addition 3 generators $c_ev_e, c_e\rho_e,\lambda_e$. Factorize the edge-corresponding $X_e$ as $u_e | c_e v_e,u_e | c_e \rho_e, \lambda_e | c_e v_e, \lambda_e | c_e \rho_e$.
    
    Otherwise, $v_e\in U$, then select in addition 3 generators $u_ec_e,\lambda_ec_e,\rho_e$. Factorize the edge-corresponding $X_e$ as $u_e c_e | v_e,u_e c_e | \rho_e, \lambda_e c_e | v_e, \lambda_e c_e | \rho_e$. The total generating set size is $3m + |U|$, and the total number of factors is $8m$.

    To prove 
    $\operatorname{rk}(X_G)\geq 3m+|U|$, take any generating set $Y$ and remove any generators in it that is not used. Let
    $$U=\{u\in Q|\text{the corresponding singleton letter $u$ belongs to $Y$} \}.$$
    Note all other generators that are not in $U$ are edge specific by definition. Then we can partition the rest of the generators according to their corresponding edge. Let $Y_e$ be the set containing the edge specific generators of edge $e$. Note
    $$|Y|=|U|+\sum_{e\in E}|Y_e|$$
    For an edge $e$, only $Y_e$ and the selected endpoint singleton letters can help generate the words of $X_e$. Lemma~\ref{lem:mingen} gives that $|Y_e|\geq 3$ for every edge, and $|Y_e|\geq 4$ when neither endpoint belongs to $U$.

    Let $a(U)$ count the edges with no endpoint in $U$. Then
    $$|Y|\geq |U|+3m+a(U).$$
    Adding one endpoint for every uncovered edge gives a vertex cover of size $|U|+a(U)$. Hence $|U|+a(U) \geq \tau(G)$, proving $|Y|\geq \tau(G)+3m.$ Together with the upper bound this proves the lemma.
    
\end{proof}

\section{Proof of Proposition~\ref{prop:gap}}
\label{pf:gap}
In this section, we'll decode the encoded corpus to obtain the gap shown in Proposition~\ref{prop:gap}. First, let's prove a simple lemma that any block of $B_{a,j}$ is recognizable in any concatenation of binary blocks. 

\begin{lemma}
    \label{lem:regBlock}
    In a concatenation of \hyperlink{bb}{binary blocks} $B_{a_0,j_0}B_{a_1,j_1}...B_{a_y,j_y}$, the occurrence of $11$ marks the beginning of each block. Moreover, any occurrence of some binary block $B_{a,j}$ in the concatenation must be one of the $B_{a_i,j_i}$'s in the concatenation
    . Any occurrence $B_{a,j}$ is uniquely identifiable to have label $(a,j)$
\end{lemma}

\begin{proof}
    After the initial 11, the remaining binary string in a binary block has a 0 in every other place, and it starts and ends with a 0, so there are no occurrence of 11 in the remaining binary string. At the boundary of binary blocks, there's a 0 from the preceding block and a 110 from the subsequent block, the only occurrence of 11 is at the beginning of the subsequent block. So the boundary does not produce any illegal occurrence of 11 either.

    Every binary block begins with the 11 marker and has length $l$. Any full occurrence of a binary block therefore aligns with one of the concatenated blocks.

    The block $B_{a,j}$ is uniquely identifiable because when we encoded the letter-phase pair $(a,j)$, we chose an \hyperlink{inj}{injective function}.
\end{proof}

Now we specify the method of decoding. 

The following lemma translates a vocabulary of a Tokenization corpus to a generating set of its corresponding source corpus.

\begin{lemma}[Tok to Fac]
    \label{tok2fac}
    Let $X$ be a corpus of $d$ non-empty source words of a FAC instance. Let \hyperlink{hG}{$h_R$}$(X)$ be the encoded corpus of the corresponding Tok instance. Let $V$ be a vocabulary for the Tok instance. Suppose the segmentation of the Tok corpus with vocabulary $V$ only uses $T$ tokens in total. If 
    \begin{align}
        T<R+d,
    \end{align} 
    then there exists some generating set $Y\subset A^+$ with $|Y|\leq |V|$. Consequently, $rk(X)\leq |V|$.
\end{lemma}

\begin{proof}
    Note we call the binary encoding of a source word a "document" in the Tok corpus.
    If document $i$ in the Tok corpus is tokenized to $t_i$ tokens, then it has $t_i-1$ internal token boundaries. So the total number of   internal boundaries in the whole corpus will be 
    \begin{align}
        \sum_{i=1}^d (t_i-1)=T-d.
    \end{align}
    Call a phase $j$ bad if any internal token boundary lies inside any block $B_{*,j}$ in any document. One internal boundary can only lie inside at most one block. If it coincides with a block boundary, it does not lie inside any block. Any one internal token boundary can mark at most one phase to be bad. So there are at most $T-d<R$ bad phases. 
    
    Recall there are $R$ phases. So we can choose a good phase $j$. A good phase is one that is not bad. From now on, we will only look at the phase $j$ parts of the Tok corpus. Delete from the corpus and vocabulary anything that is not phase $j$. This is achievable because Lemma~\ref{lem:regBlock} tells us the binary blocks are uniquely recognizable. So we can recognize the blocks in the corpus and identify which blocks are phase $j$.
    
     From the updated corpus, we can still recover the whole source corpus $X$ because the deleted phases are merely repeats of the good phase $j$. Moreover, we can derive a factorization of $X$ from the tokenization of the corpus. For each source word $x\in X$, suppose its phase-$j$-tokenization is as follows
    \begin{align}
        h_R(x) = t_1...t_s
    \end{align} 
    where $t_i$ are tokens. Since our chosen phase $j$ is good, each $t_i$ is consist of complete phase-$j$-binary-blocks because no token boundary lies inside a binary block. By Lemma~\ref{lem:regBlock}, we can recognize these binary blocks. Say we recognize the above document $h_R(x)$ to have blocks 
    \begin{align}
        B_{a_0,j},B_{a_1,j},...B_{a_\nu,j},
    \end{align}
    then we recover the source word $x$ to be
    \begin{align}
        x=a_0a_1...a_\nu
    \end{align}
    Note $s\leq \nu$ because one token $t_i$ could contain multiple binary blocks.
     In addition, separating the word $x$ at the corresponding token boundaries in $h_R(x)$ provide a \textit{factorization} of $x$.

     Now we say what the generating set should be. Say a phase-$j$ token $t$ contains binary blocks 
     $$t=B_{a_0,j},B_{a_1,j},...B_{a_\xi,j}.$$
     We want to "read off" the corresponding word that was mapped to this token $t$. So we define the translation function $\pi(t):=a_0a_1...a_\xi$. This is doable because, for each binary block in token $t$, one can first take its inverse of the one-to-one encoding function \hyperlink{z}{$z$}. Then one can output the first coordinate in that inverse.
     Let
     \begin{align}
         Y := \{\pi(t) | t\in V \}
     \end{align}
    The map $\pi$ is a function of the source-letter-spelling of its input token $t$, so there is at most one image for each vocabulary entry. Hence, $|Y| \leq |V|$. By the reasoning above, we know that $Y$ factorizes $X$, so it is a generating set of $X$.
    
\end{proof}

\textbf{Proposition~\ref{prop:gap}}
\textit{
For every integer $R \ge 1$, \hyperref[Transformation1]{Transformation~\ref*{Transformation1}} satisfies
 \begin{align}
rk(X_G) \le B
&\implies \operatorname{OPT}_K(D_R) \le 8m,\\
rk(X_G) > B
&\implies \operatorname{OPT}_K(D_R) \ge R+4m.
\end{align} }

\begin{proof}
    Recall in \hyperref[Transformation1]{Transformation~\ref*{Transformation1}}, we set $K=B$.
    If $\operatorname{rk}(X_G)\leq B = 3m+\Phi $, then the proof of Lemma~\ref{lem:source-rank} describes a generating set $Y$ of at most $3m + \Phi = B$ words and exactly $8m$ factors in its factorization of $X_G$. We can simply set $V:=\{h_R(y)|y\in Y\}$, and $|V|\leq |Y|\leq B=K$. Then we tokenize the Tok corpus the same way as the factorization of $X_G$, substituting each factor $y\in Y$ with the corresponding token $h_R(y)$. Then we get a segmentation length of the Tok corpus of length $8m$.

    Suppose $\operatorname{rk}(X_G)> B=K$. Recall $\hyperlink{d}{d=4m}$. If some vocabulary $V$ of cardinality at most $K$ had a tokenization with fewer than $R + 4m = R + d$ tokens, then Lemma~\ref{tok2fac} would produce a generating set of size at most $|V|\leq K=B$, a contradiction. The lemma applies to every feasible vocabulary and every tokenization, including minimum-token ones. 
\end{proof}

\section{Proof of Theorem~\ref{thm:storage-bicriteria}} \label{pf:storage}
\begin{proof}
Restrict the candidate words to those of length at most $B$,
and replace the bundle cost by
\[
    c(Q_w)=\lambda |w|+|Q_w|.
\]
Every vocabulary of storage cost at most $B$ uses only such
words. The dynamic program in Algorithm~\ref{algo:FWB}
remains unchanged, except that the density for a bundle
of size $j$ is now
\[
    \frac{F_w(m_w,j)}{\lambda |w|+j}.
\]
Singletons retain cost $1$.

Let $V^*$ be an optimal vocabulary with $c(V^*)\leq B$.
As in Lemma~\ref{lem:ucb-segmentation}, its induced
segmentation cover has cost at most
\[
    \lambda c(V^*)+\OPT^{st}_B
    \leq \lambda B+\OPT^{st}_B.
\]
Thus, the charging argument of
Lemma~\ref{lem:ucb-greedy} gives
\[
    \lambda A+R
    \leq \alpha_n(\lambda B+\OPT^{st}_B),
    \qquad
    \alpha_n=1+\ln n,
\]
where $A=c(V_g)$ and $R$ is the total number of selected
occurrences, including singletons. Here, each distinct
word in $V_g$ contributes its storage cost at least once
to the selected bundles.

We next replace the chunk construction by a binary
decomposition. For each $w\in V_g$, recursively split
$w$ into two consecutive parts whose lengths differ
by at most one, stopping at single characters.
Let $V_w$ be the set of distinct words of length at
least two represented by nodes of this decomposition,
including the root $w$.

The decomposition has height at most
$\lceil\log_2|w|\rceil\leq D_B$.
At each depth, the represented substrings are disjoint
parts of $w$, so their total length is at most $|w|$.
Since only the first $D_B$ depths can contain nonleaf
nodes,
\[
    c(V_w)\leq D_B|w|.
\]
Moreover, every nonempty suffix of $w$ can be segmented
into at most $D_B$ tokens from $V_w\cup\Sigma$.
To see this, follow the path to the first character
of the suffix. Whenever this path enters a left child,
include the entire right child. Together with the
first character, these nodes partition the suffix.
The two bottom leaves, if both included, can be
replaced by their parent. Consequently, a path of
length at most $D_B$ contributes at most $D_B$ tokens.

Set
\[
    V=\bigcup_{w\in V_g}V_w.
\]
Then
\[
    c(V)\leq\sum_{w\in V_g}c(V_w)\leq D_B A.
\]
Applying the same interval-selection procedure as in
Lemma~\ref{lem:ucb-partition}, with the suffix
decomposition above in place of the chunk decomposition,
gives a valid segmentation using at most $D_B R$ tokens.

Finally, consider the guesses $t=1,\ldots,n$ in increasing
order, set $\lambda=t/(\eta B)$, and accept the first
cover satisfying
\[
    \lambda A+R\leq\alpha_n t(1+1/\eta).
\]
The greedy bound ensures that this condition holds
at $t=\OPT^{st}_B$. Hence, the first accepted guess
$\hat t$ satisfies $\hat t\leq\OPT^{st}_B$, and
\[
    A\leq(1+\eta)\alpha_n B,
    \qquad
    R\leq(1+1/\eta)\alpha_n\OPT^{st}_B.
\]
Combining these inequalities with the bounds above gives
\[
    c(V)\leq(1+\eta)(1+\ln n)D_B B,
    \qquad
    T(\calD,V)\leq
    (1+1/\eta)(1+\ln n)D_B\OPT^{st}_B.
\]

The modified bundle computation and the enumeration of
guesses run in polynomial time by the same runtime
analysis as Algorithm~\ref{Final_algo}. Each binary
decomposition has $O(|w|)$ nodes and can also be
constructed in polynomial time. Fixing a deterministic
rule for ties makes the algorithm deterministic.
Taking $\eta=1$ gives equal factors
$2(1+\ln n)\lceil\log_2 B\rceil=O(\log n\log B)$.
\end{proof}

%% file: sections/index.tex
\clearpage
\section*{Index of Notation}
\addcontentsline{toc}{section}{Index of Notation}

\begingroup
\small

\newcommand{\BHPVnotation}[2]{%
  \par\noindent\hangindent=1.5em\hangafter=1
  \(#1\)\enspace #2\par
}

\subsection*{Corpus and tokenization}

\BHPVnotation{\Sigma,\ c,\ \epsilon}
  {The binary alphabet \(\Sigma=\{0,1\}\), a character \(c\in\Sigma\),
   and the empty word \(\epsilon\).}

\BHPVnotation{A^*,\ A^+,\ \Sigma^{>1}}
  {The set of finite words over \(A\), the set of nonempty words over
   \(A\), and the set of binary words of length at least two, respectively.}

\BHPVnotation{\mathcal D,\ d,\ s_i}
  {The corpus, the number of documents, and the \(i\)-th document.
   When a document \(s\) is fixed, \(s_i\) also denotes its \(i\)-th
   character.}

\BHPVnotation{n,\ M}
  {The total corpus length \(n=\sum_{i=1}^{d}|s_i|\) and the maximum
   document length \(M=\max_{1\le i\le d}|s_i|\).}

\BHPVnotation{\mathcal W(\mathcal D),\ \mathcal W,\ w,\ V}
  {The set of distinct substrings of length at least two occurring
   within a document of \(\mathcal D\), abbreviated \(\mathcal W\).
   A candidate word is denoted by \(w\), and a selected vocabulary
   by \(V\subseteq\mathcal W\).}

\BHPVnotation{w_1,\ldots,w_t}
  {The tokens of a segmentation \(s=w_1\cdots w_t\), where
   \(w_j\in V\cup\Sigma\) and \(t\) is its token count.}

\BHPVnotation{\mathrm{Tok}(s,V),\ T(\mathcal D,V)}
  {The minimum token count for \(s\) over \(V\cup\Sigma\), and the
   minimum total token count
   \(T(\mathcal D,V)=\sum_{i=1}^{d}\mathrm{Tok}(s_i,V)\).}

\BHPVnotation{K,\ \mathrm{OPT}_K(\mathcal D)}
  {The vocabulary cardinality budget and the minimum of
   \(T(\mathcal D,V)\) over all \(V\) with \(|V|\le K\).
   We write \(\mathrm{OPT}_K\) when the corpus is understood.}

\BHPVnotation{B,\ c(V),\ \mathrm{OPT}^{\mathrm{st}}_B(\mathcal D)}
  {The storage budget, the spelling cost \(c(V)=\sum_{w\in V}|w|\),
   and the minimum of \(T(\mathcal D,V)\) over all \(V\) with
   \(c(V)\le B\).}

\BHPVnotation{\alpha,\ \beta}
  {The budget-augmentation factor and the token-count approximation
   factor, respectively.}

\BHPVnotation{H_n,\ \alpha_n}
  {The harmonic number \(H_n=\sum_{j=1}^{n}1/j\) and its upper bound
   \(\alpha_n=1+\ln n\).}

\BHPVnotation{q}
  {In the single-string storage bound, \(q=\min\{B,n-B\}\).}

\subsection*{Vertex cover and word factorization}

\BHPVnotation{G=(Q,E),\ w,\ m}
  {The source graph, with vertex set \(Q\), edge set \(E\),
   vertex count \(w=|Q|\), and edge count \(m=|E|\).}

\BHPVnotation{\tau(G),\ \Phi}
  {The minimum size of a vertex cover of \(G\) and the vertex-cover
   decision budget.}

\BHPVnotation{A,\ X,\ Y,\ Y^*,\ \mathrm{rk}(X)}
  {The source alphabet, a finite set of source words, a generating
   set, the set of finite concatenations of elements of \(Y\), and
   the minimum size of a set \(Y\) satisfying \(X\subseteq Y^*\).}

\BHPVnotation{e=(u_e,v_e),\ c_e,\ \lambda_e,\ \rho_e}
  {The endpoints of edge \(e\), also used as singleton letters,
   and the three letters private to \(e\). Lemma A.1 writes these
   five letters as \(u,v,c,\lambda,\rho\).}

\BHPVnotation{X_e,\ X_G}
  {The four source words associated with edge \(e\), and the full
   source corpus \(X_G=\bigcup_{e\in E}X_e\). Specifically,
   \(X_e=\{u_ec_ev_e,\ u_ec_e\rho_e,\ \lambda_ec_ev_e,\
   \lambda_ec_e\rho_e\}\).}

\BHPVnotation{N,\ B}
  {The total source length \(N=\sum_{x\in X_G}|x|=12m\) and the
   generator budget \(B=3m+\Phi\). The reduction sets \(K=B\).}

\BHPVnotation{U,\ Y_e,\ a(U)}
  {In the rank proof, \(U\subseteq Q\) consists of vertices whose
   singleton letters belong to \(Y\), \(Y_e\) consists of generators
   private to \(e\), and \(a(U)\) counts edges with neither endpoint
   in \(U\).}

\subsection*{Binary encoding and decoding}

\BHPVnotation{\varepsilon,\ q,\ p,\ R}
  {The hardness exponent \(0<\varepsilon<1\), the amplification
   exponent \(q=\lceil4/\varepsilon\rceil\), the base \(p=N+2q+10\),
   and the number of phases \(R=p^q\). Proposition 3.2 allows
   any integer \(R\ge 1\).}

\BHPVnotation{a,\ j,\ b,\ z(a,j)}
  {A source letter, a phase \(j\in\{1,\ldots,R\}\), the bit width
   of a label, and the distinct \(b\)-bit label assigned to
   the pair \((a,j)\).}

\BHPVnotation{z,\ z_i,\ C(z),\ B_{a,j},\ l}
  {A label \(z=z_1\cdots z_b\), its \(i\)-th bit, its padded
   encoding \(C(z)=110z_1 0z_2 0\cdots z_b0\), the binary block
   \(B_{a,j}=C(z(a,j))\), and the block length \(l=2b+3\).}

\BHPVnotation{x=a_1\cdots a_r,\ h_R}
  {A source word of length \(r\) and its encoding map.
   Each \(a\) is replaced by \(h_R(a)=B_{a,1}\cdots B_{a,R}\),
   and \(h_R(x)=h_R(a_1)\cdots h_R(a_r)\).}

\BHPVnotation{h_R(X),\ D_R}
  {The encoded corpus of a source set \(X\), and the particular
   corpus \(D_R=h_R(X_G)\). It has \(d=4m\) documents and total
   length \(n=NRl\).}

\BHPVnotation{T,\ t_i,\ B_{*,j}}
  {In the boundary count, \(T\) is the number of tokens in the
   encoded corpus, \(t_i\) is the number in document \(i\), and
   \(B_{*,j}\) denotes a phase-\(j\) block with arbitrary source
   letter. The total number of internal token boundaries is \(T-d\).}

\BHPVnotation{y,\ \nu,\ \xi}
  {In Appendix B, the final zero-based block indices in a
   concatenation of blocks, a projected document, and a projected
   token, respectively. In the proof of Proposition 3.2,
   \(y\in Y\) denotes a generator.}

\BHPVnotation{t_1,\ldots,t_s,\ \pi(t)}
  {In the decoding argument, \(t_1,\ldots,t_s\) are the tokens in
   the displayed phase projection. The map \(\pi(t)\) reads the
   source letters of the retained blocks of token \(t\), in order.}

\subsection*{Occurrences, bundles, and greedy covering}

\BHPVnotation{I=(s,[i,j))}
  {An occurrence covering positions \(i,\ldots,j-1\) of document
   \(s\). Subscripts on \(I,i,j,s\) distinguish occurrences.}

\BHPVnotation{\mathcal O_w,\ m_w}
  {The occurrences of \(w\) in \(s\), all its occurrences in the
   corpus, and their number \(m_w=|\mathcal O_w|\).}

\BHPVnotation{Q_w,\ Q,\ \lambda,\ c(Q_w)}
  {A nonempty bundle of pairwise disjoint occurrences of \(w\),
   also written \(Q\), its opening charge \(\lambda>0\), and its
   cardinality-model cost \(c(Q_w)=\lambda+|Q_w|\).}

\BHPVnotation{\mathcal P,\ \mathcal B,\ S}
  {A segmentation and the two parts of its induced cover:
   \(\mathcal B\) is the family of nonempty word bundles and \(S\)
   is the set of positions covered by singleton tokens.
   The induced cover is also written \(\mathcal P=(\mathcal B,S)\).}

\BHPVnotation{V',\ k,\ t,\ c(\mathcal P)}
  {The set of vocabulary words used by the segmentation, its size
   \(k=|V'|\), the token count \(t\), and the induced cover cost
   \(c(\mathcal P)=\lambda k+t\).}

\BHPVnotation{\mathbf D,\ I_i,\ a_i,\ b_i,\ p_i}
  {The concatenation \(\mathbf D=s_1\cdots s_d\) used to index
   positions, the \(i\)-th occurrence of a fixed word in start
   order, its interval \(I_i=[a_i,b_i)\), and the last index
   \(k<i\) with \(b_k\le a_i\). If no such \(k\) exists, \(p_i=0\).}

\BHPVnotation{U,\ g_i}
  {The currently uncovered positions and the gain
   \(g_i=|U\cap I_i|\) from selecting occurrence \(I_i\).}

\BHPVnotation{F_w(i,j)}
  {The maximum number of newly covered positions obtainable by
   selecting exactly \(j\) disjoint occurrences among
   \(I_1,\ldots,I_i\) of word \(w\).}

\BHPVnotation{\mathcal O,\ G,\ P,\ F,\ m}
  {In Algorithms 1--3, the occurrence lists, gain arrays,
   predecessor arrays, dynamic-programming table, and number of
   occurrences of the current word. For fixed \(w\), the entries
   \(\mathcal O[w][i]\), \(G[w][i]\), and \(P[w][i]\)
   correspond to \(I_i\), \(g_i\), and \(p_i\), respectively.}

\BHPVnotation{\rho,\ \rho^*,\ j^*,\ Q^*,\ w^*,\ \bot}
  {The candidate density (newly covered positions per unit cost),
   best density, maximizing bundle size, best bundle, and its word.
   The sentinel \(w^*=\bot\) selects the singleton alternative.}

\BHPVnotation{x,\ J}
  {In Algorithm 2, the position chosen for a singleton and the set
   of positions newly covered in the current iteration.}

\BHPVnotation{\mathcal C,\ \mathcal C_{\mathcal W},\
              \mathcal C_\Sigma,\ R}
  {The greedy cover, its selected bundles, its selected singletons,
   and the total number of selected occurrences, including
   singletons:
   \(R=\sum_{Q\in\mathcal C_{\mathcal W}}|Q|+|\mathcal C_\Sigma|\).}

\BHPVnotation{V^*,\ C,\ c(\mathcal C)}
  {An optimal vocabulary for budget \(K\), the cost of its induced
   comparison cover, and the cost of the greedy cover, respectively.}

\BHPVnotation{r,\ r_i,\ z_i,\ C_i,\ L}
  {In the greedy cost proof, \(r\) is the number of uncovered
   positions. At iteration \(i\), \(r_i\) is the number remaining
   before selection, \(z_i\) is the number newly covered, and
   \(C_i\) is the selection cost. The algorithm takes \(L\) iterations.}

\subsection*{Substring chunks and budget augmentation}

\BHPVnotation{\mathrm{Sub}(w),\ \mathrm{Suf}(w)}
  {The substrings of \(w\) of length at least two and the nonempty
   suffixes of \(w\), respectively. In \(w=xuy\) or \(w=xu\),
   \(u\) is the indicated substring or suffix.}

\BHPVnotation{q,\ \eta,\ h,\ \gamma_q}
  {The maximum number of chunks per selected word, the positive
   rational tradeoff parameter, the suffix token bound
   \(h=q-1+\lceil M/q\rceil\), and the common bicriteria factor
   \(\gamma_q=(q+h)(1+\ln n)\) obtained when \(\eta=h/q\).}

\BHPVnotation{l,\ r,\ b,\ V_w,\ y}
  {In the chunk construction, \(l=|w|\), \(r=\lceil l/q\rceil\)
   is the chunk length, \(b=\lceil l/r\rceil\) is the number of
   chunks, \(V_w\) is the set of distinct chunks of length at least
   two, and \(y\) is a nonempty suffix of \(w\).}

\BHPVnotation{x,\ a,\ b}
  {In the cover-to-partition proof, \(x\) is the first unprocessed
   position and \([a,b)\) is a selected interval containing \(x\).
   In Algorithm 3, \(a,b\) are the endpoints of the current chunk.}

\BHPVnotation{t,\ \hat t,\ \lambda}
  {In Algorithm 3, \(t\) is the trial token count, \(\hat t\) is the
   first accepted trial, and \(\lambda=t/(K\eta)\). At output,
   \(t\) denotes the token count of the returned segmentation.}

\BHPVnotation{V_g,\ A,\ V,\ \mathcal P}
  {The set of distinct words selected by the greedy cover, its
   cardinality \(A=|V_g|\), the final chunk vocabulary
   \(V=\bigcup_{w\in V_g}V_w\), and the output segmentation.}

\subsection*{Storage-budget cover}

\BHPVnotation{D_B,\ \mathcal W_B,\ \mathrm{OPT}_B}
  {The parameter \(D_B=\lceil\log_2 B\rceil\), the candidate set
   \(\mathcal W_B=\{w\in\mathcal W:|w|\le B\}\), and the shorthand
   used in Appendix C for the storage optimum
   \(\mathrm{OPT}^{\mathrm{st}}_B\).}

\BHPVnotation{\lambda,\ c(Q_w)}
  {The per-unit spelling charge \(\lambda=t/(\eta B)\) for trial
   \(t\), and the bundle cost \(c(Q_w)=\lambda|w|+|Q_w|\).}

\BHPVnotation{\mathcal C_{\mathcal W_B},\ L,\ R}
  {The bundles selected in the storage model, their total spelling
   length \(L=\sum_{Q_w\in\mathcal C_{\mathcal W_B}}|w|\), and the
   number of selected occurrences including singletons. A spelling
   contributes once per selected bundle, so the cover costs
   \(\lambda L+R\).}

\endgroup